\documentclass[12pt]{article}

\def\fnote#1#2{\begingroup\def\thefootnote{#1}\footnote{#2}
    \addtocounter{footnote}{-1}\endgroup}

\newcommand{\be}{\begin{equation}}
\newcommand{\ee}{\end{equation}}
\newcommand{\bea}{\begin{eqnarray}}
\newcommand{\eea}{\end{eqnarray}}

\def\abstract#1{\begin{center}{\large ABSTRACT}\end{center} \par #1}
\def\title#1{\begin{center}{\Large\bf {#1}}\end{center}}
\def\author#1{\begin{center}{\large #1}\end{center}}
\def\address#1{\begin{center}{\it #1}\end{center}}

\begin{document}
\begin{titlepage}
\hspace*{\fill}
\vbox{
\hbox{UT-948}
\hbox{hep-th/0106253}
}
\vspace*{\fill}
\begin{center}
\Large\bf
Non-commutative branes in D-brane backgrounds
\end{center}
\vskip 1cm
\author{
Masako Asano \fnote{\dag}{ 
E-mail address: {\tt asano@hep-th.phys.s.u-tokyo.ac.jp}}}
\address{
Department of Physics, \\
  University of Tokyo, \\
 Hongo, Bunkyo-ku, Tokyo 113-0033 Japan}
\vspace{\fill}
\abstract{
We study Myers world-volume effective action of coincident D-branes. 
We investigate a system of $N_0$ D0-branes in the geometry of 
D$p$-branes with $p=2$ or $p=4$.
The choice of coordinates can make the action simplified and tractable.
For $p=4$, we show that a certain point-like D0-brane configuration
solving equations of motion of the action
can expand to form a fuzzy two-sphere 
via magnetic moment effect without changing quantum numbers.
We compare non-commutative D0-brane configurations 
with dual spherical D$(6-p)$-brane systems.
We also discuss 
the relation between these configurations and giant gravitons in 11-dimensions.
}

\vspace*{\fill}



\end{titlepage}

\section{Introduction}
Effective action of $N_p$ coincident D$p$-branes 
has been extended to include couplings to Ramond-Ramond field strength $F^{(n)}$
of any form degree $n$ other than that of $n\le p+2$ \cite{My, TR}.
The couplings of the D$p$-branes with $F^{(n)}$ ($n\ge p+4$)
induce dielectric effects (Myers effects) \cite{My}
or their magnetic analogues 
(magnetic moment effects) \cite{MST,DTV}.
The effects are important in understanding the nature of non-abelian 
D-brane systems and  have been studied in various contexts 
\cite{My,CMT,Hy,TV}.
In particular, they play an important role \cite{PS, Be, BN, Za, AI}
in the context of AdS/CFT correspondence \cite{AGMOO}.

Most analyses given so far concern flat spacetime where the RR field strength $F^{(n)}$ is put
as an external field \cite{My,CMT,Hy,TV}.
Such backgrounds ignore the back reaction of $F^{(n)}$, i.e.,
they do not solve supergravity equations of motion.
For example, consider the effective action of D0-branes in
the background with flat metric and $F^{(4)}_{\mu 123}\!=\!const.$ \cite{My,DTV}.
For $\mu=0$, a static non-commutative configuration of D0-branes 
other than the usual point-like configuration can 
solve equations of motion of the action due to 
the dielectric effect \cite{My}.
Similarly for $\mu=4$, D0-branes can expand via 
magnetic moment effect \cite{DTV}
and there are a point-like and a non-commutative
configuration with a constant momentum $P_4$ 
as solutions of the action.
In each of these cases, the energy of 
the non-commutative configuration of $N_0$ D0-branes
is less than that of corresponding commutative configuration.
This means that the true ground state of the model is given by the configuration 
of expanding D0-branes.
On the other hand, if we consider to put $F^{(6)}_{\mu 12345}\!=\!const.$
in flat background, a non-commutative configurations of D0-branes is 
unstable and a stable configuration is given by point-like D0-branes \cite{TV}.

If we deal with a supergravity background with some remaining supersymmetries, 
we expect to have a BPS configuration of commutative D0-branes.
Thus if there exists an expanding configuration with the same number of supersymmetries, 
it would have the same energy as the commutative configuration.
One of our purposes of this paper 
is to see if such a structure certainly exists in D$p$-brane backgrounds.
Note that ref.\cite{DTV} gives similar discussion in the near horizon geometry
of D$4$-branes.
There it was shown that 
substitution of certain non-commutative matrices $X^i$ to the 
non-abelian action of D0-branes yields 
the same terms as the expansion of the dual spherical D2-brane action.
We shall try to argue the problem from the microscopic point of view 
by explicitly solving the equations of motion of D0-branes
in D$p$-brane backgrounds.
We consider the coupling of D0-branes with the magnetic RR field strength
$F^{(8-p)}$ associated with the D$p$-branes.
Thus we only deal with magnetic moment effect.

One of the difficulties of dealing with the action in general backgrounds is 
that it is coordinate dependent
due to the appearance of commutators of U($N_0$) adjoint scalars $X^i$ and 
the non-abelian Taylor expansions. 
We shall see that in some cases we can avoid the difficulties  
and give explicit form of the action by choosing coordinate system properly.

We explicitly consider the geometry 
of D$p$-branes with $p=2$ or $p=4$ and analyze the magnetic moment effect.
For such a background, we show that we can choose appropriate coordinate system 
and can write the non-abelian action of D0-branes explicitly.
In particular, for D4-brane background, 
we find two solutions with the same energy and momentum
by solving equations of motion of the action explicitly if we take 
near horizon limit.
One of them is point-like and the other is fuzzy spherical.
We also see that each of these systems is related to 
the spherical D$(6-p)$-branes with an appropriate U(1) field $F_{ab}$ on the
branes.

\bigskip

This paper is organized as follows.
In section 2, we briefly review Myers non-abelian action of D-branes and consider
the application to general backgrounds.
In section 3, we analyze the action in the geometry of D$p$-branes
with $p=2$ or $p=4$ and try to solve the equations of motion.
We compare the non-abelian action with that of spherical D$(6-p)$-brane action.
Section 4 gives summary and discussion.
In particular, we briefly discuss the relation between expanding D0-brane 
via Myers effects and rotating spherical M-branes 
(giant gravitons) in $AdS_m \times S^n$ in 11-dimensions.

\section{Effective action of coincident D-branes}
The world-volume effective action of $N_p$ coincident D$p$-branes 
in type IIA or IIB theory is given by Myers \cite{My}.
It is constructed from the D9-brane action in type IIB theory
by applying T-duality transformation along $9-p$ space coordinates 
$x^{p+1}\!,\cdots ,x^{9}$.
The action involves a U($N_p$) gauge field $A_a$ 
and $9-p$ adjoint scalars $X^i$ (i=$p+1,\cdots ,9$)
(and their superpartners).
The field strength of 
$A_a$ is $F_{ab}=\partial_a A_b -\partial_b A_a + i [A_a, A_b]$ and 
the covariant derivative of $X^i$ is 
$D_a X^i =\partial_a X^i+i [A_a, X^i]$ 
as usual.
We choose static gauge $x^a=\sigma^a$ ($a=0,1,\cdots,p$)
for spacetime coordinates $x^\mu$
and worldvolume coordinates $\sigma^a$.
Suppose that Dp-branes are put in the background of string frame metric 
$ds^2=G_{\mu\nu} dx^\mu dx^\nu $,
NS-NS 2-form potential $B_{\mu\nu}$ and $n$-form RR potentials 
$C^{(n)}$ for $n=\cdots ,p-1,p+1,\cdots$.
Then the proposed action is written as sum of the
Born-Infeld action $S_{BI}$ and the Chern-Simons action $S_{CS}$.
Bosonic part of $S_{BI}$ is given as
\be
S_{BI}=-T_p\int d^{p+1}\sigma
{\rm Tr} \left[
e^{-\phi} \sqrt{
-\det(P[E_{ab}+E_{ai}(Q^{-1}-\delta)^{i}{}_{k}E^{kj}E_{jb}]+\lambda F_{ab})
\det(Q^i{}_j)
}
\right]
\label{MpBI}
\ee
where $\lambda=2\pi l_s^2$, 
$E_{\mu\nu}=G_{\mu\nu}+B_{\mu\nu}$, $T_{p}=2\pi/g_s(2\pi l_s)^{p+1}$
is the tension of the D$p$-brane and 
\be
Q^i{}_j=\delta^i{}_j+i \lambda^{-1} [X^i,X^k]E_{kj}.
\ee
The pull-back $P[\cdots]$ is defined as
\be
P[Z_{a_1\cdots a_n}]=Z_{\mu_1\cdots \mu_n}
D_{a_1}X^{\mu_1}\cdots D_{a_n} X^{\mu_n}\,.
\ee 
The field $G_{\mu\nu}$ (or $B_{\mu\nu}$, $\phi$) 
in the above action is a functional of $X^{i}$ and the explicit form would
be given by a non-abelian Taylor expansion 
of the corresponding background fields
as \cite{GM}, e.g., 
\be
G_{\mu\nu}(X^i(\sigma^a),\sigma^a) =
\sum^\infty_{n=0} \frac{1}{n!} X^{i_1}\cdots X^{i_n} 
\partial_{i_1}\cdots \partial_{i_n}G_{\mu\nu}(x^i,\sigma^a)|_{x^i=0} \,.
\ee
In the above expression we assume that the
D$p$-brane is temporarily put on $x^i=0$ and the scalar fields $X^i$ represent
fluctuation around $x^i=0$.  
Chern-Simons action is given as 
\be
S_{CS}=T_p \int {\rm Tr}\left(
P\left[e^{i\lambda^{-1}i_{X}i_{X}}(\sum C^{(n)}e^B)
\right] e^{\lambda F}
\right) .
\label{MpCS}
\ee
Here $i_X$ is an interior product which reduces the form degree $-1$
as e.g.,
\be
i_Xi_X C^{(2)}=\frac 1 2 C_{ij}^{(2)}[X^j, X^i].
\ee
This interior product induces the coupling of the D-branes to
the RR potential of higher degree $n=p+3,p+5,\cdots$.
Note that we interpret Tr( ) as symmetrized trace:
we take the traces after we symmetrize 
 all $F_{ab}$, $D_a X^i$, $[X^i,X^j]$ and each $X^i$ appearing
in the non-abelian Taylor expansions.

In practice, the action eq.(\ref{MpBI}) or eq.(\ref{MpCS})
can be justified a priori only when there is an isometry along 
each $x^i$ since it is constructed from the D9-brane action 
via T-duality transformation \cite{My}.
However, as was described above,
the action can formally be defined for any coordinate system 
$\{x^\mu\}$ in any spacetime if we use non-abelian Taylor expansion 
of background fields $E_{\mu\nu}$, $\phi$ and $C^{(n)}$.
For such a non-trivial background, the action does not have 
general covariance anymore and the meaning of the action is not clear enough.
In spite of the subtlety, we would like to consider 
the action of a set of coincident D$p$-branes in
a particular background without an isometry along $x^i$ 
($i=p+1,\cdots,9$).
We have to define $G_{\mu\nu}$, $B_{\mu\nu}$ and $C^{(n)}$ by using 
non-abelian Taylor expansion as described above
if they are not constants.
Thus in general, the action can describe the 
behavior of the D-branes only around $x^i \sim 0$.
It is not suitable enough to deal with expanding brane configurations.

We would like to determine the Chern-Simons action without using infinite 
series of Taylor expansion since it is important to 
obtain expanding brane configurations.
We seek for a coordinate system 
where the field $C^{(n)}$ becomes independent of $X^i$ or 
is represented as a finite polynomial of $X^i$.
Even if there is no such coordinate system, 
we can consider a model that some of the adjoint scalar fields 
$X^i$ are set to be diagonal from the beginning: $X^i= x^i{\bf 1}$. 
In some cases, such a model does not need to use non-abelian
Taylor expansion
if we choose an appropriate coordinate system.
In practice, we use such method to analyze the action of D0-branes in the 
D$p$-brane background in the next section.

Now we especially consider the action of $N_0$ D0-branes for our future 
purpose.
Each term of the Chern-Simons action can be expanded by polynomials of $X^i$
around $X^i=0$.
For example, couplings of D0-branes with the background $C^{(3)}$ and 
$C^{(5)}$ in ${\cal L}_{CS}$ is respectively given as
\bea
{\cal L}^{3}_{CS} &=& i \frac{T_0}{\lambda} {\rm Tr} P[(i_X i_X) C^{(3)}]
\nonumber\\
&=& i \frac{T_0}{2 \lambda}{\rm Tr}\left(
C^{(3)}_{tij}(X,\sigma)[X^j,X^i]+ C^{(3)}_{ijk}(X,\sigma)[X^k,X^j]\partial_tX^i
\right)
\nonumber\\
&=&
i \frac{T_0}{3 \lambda}{\rm Tr}(X^iX^jX^k) F^{(4)}_{tijk} 
-i \frac{T_0}{4 \lambda}{\rm Tr}(X^iX^jX^k\dot{X}^l) F^{(4)}_{ijkl}
+i \frac{T_0}{4 \lambda}{\rm Tr}(X^iX^jX^kX^l) \partial_lF^{(4)}_{tijk}
\nonumber\\
&& -i \frac{T_0}{30 \lambda}{\rm Tr}(X^iX^jX^kX^l\dot{X}^m)
\left[4 \partial_{(i}F^{(4)}_{l)jkm} -2 \partial_{(j}F^{(4)}_{k)lmi}
\right]
\nonumber\\
&& -i \frac{T_0}{30 \lambda}{\rm Tr}(X^iX^jX^kX^lX^m)
\left[\partial_{i}\partial_l F^{(4)}_{tkjm} 
+2 \partial_i \partial_{j}F^{(4)}_{tklm}
\right]
 \quad + O(X^6) \, ,
\label{CS3}
\eea
\bea
{\cal L}^{5}_{CS} &=& -\frac{T_0}{2\lambda^2} {\rm Tr} P[(i_X i_X)^2 C^{(5)}]
\nonumber\\
&=& -\frac{T_0}{8 \lambda^2}{\rm Tr}\left(
C^{(5)}_{tijkl}(X,\sigma)[X^j,X^i][X^l,X^k]
+ C^{(5)}_{ijklm}(X,\sigma)[X^j,X^i][X^l,X^k]\partial_tX^m
\right)
\nonumber\\
&=&
\frac{T_0}{10 \lambda^2}{\rm Tr}(X^iX^jX^kX^lX^m) F^{(6)}_{tijklm}
-\frac{T_0}{12 \lambda^2}{\rm Tr}(X^iX^jX^kX^lX^m \dot{X}^n) F^{(6)}_{ijklmn}
\nonumber\\
&& +\frac{T_0}{12 \lambda^2}{\rm Tr}(X^iX^jX^kX^lX^mX^n) 
\partial_n F^{(6)}_{tijklm} \quad + O(X^7)
\label{CS5}
\eea
where each $F^{(n)}$ or its derivative in 
the above equations denotes the value at $X^i=0$ : e.g., $F^{(n)}|_{X^i=0}$.
Here we use the symmetrized trace prescription. 
For some other definition of trace, 
not all terms are collected in the form of field strength $F^{(n+1)}$.
In practice, gauge invariance of ${\cal L}_{CS}$ is proven 
when symmetrized trace prescription is applied \cite{Ci}.

The Born-Infeld action for coincident D0-branes reduces to 
\be
S_{BI} =-T_0\int dt {\rm Tr} \left(
e^{-\phi} \sqrt{
-[E_{00} + (Q^{-1})^j{}_k \dot{X}^i \dot{X}^k E_{ij}]
\det(Q^i{}_j)
}\right)
\ee
when we fix the gauge $E_{0i}=0$ and $A_0=0$.
The matrix ${(Q^{-1})}^i{}_j$ which is inverse to $Q^{i}{}_{j}$
is defined by polynomial expansion of $X^i$.
It is known that this form of the action can be reliable only up to
fourth order in $F_{ab}$ (or $D_a X^i$, $[X^i,X^j]$) \cite{Ts,HT}.
The problem is not critical for our analysis since
we mainly deal with up to second order of the commutators $[X^i,X^j]$.

\section{D0-branes in the geometry of D$p$-branes}
In this section we consider the non-abelian action of $N_0$ D0-branes
in supergravity backgrounds. 
Since we expect to have a configuration of branes that are
expanding into $S^2$ or $S^4$ via Myers effect,
we deal with a background which has SO$(3)$ or SO$(5)$ symmetry.
Here we consider the geometry of D$p$-branes with $p=4$ or $p=2$.

\subsection{D0-branes in the D$p$-brane geometry}
Geometry of D$p$-branes is described by
the string frame metric 
\cite{HS,St}
\bea
ds^2_{p} &=& H^{-\frac 1 2}
\eta_{\mu\nu}dx^{\mu} dx^{\nu}
+H^{\frac 1 2}\delta_{mn}dx^{m} dx^{n}
\\
e^{\phi} &=& H^{\frac{3-p}{4}}
\eea
where $r=\sqrt{x^m x_m}$, $\mu,\nu \in\{0,\cdots, p\}$, 
$m,n\in\{p+1,\cdots 9\}$ and
\be
H=1+\frac{k}{r^{7-p}}
\ee
with
\be
k=\frac{N}{7-p}\frac{2\pi}{T_{6-p}V_{8-p}}.
\label{condNk}
\ee
Here $V_q$ is the volume of unit $q$-sphere :
\be
V_q=\frac{2\pi^{\frac{q+1}{2}}}
{\Gamma \left(\frac{q+1}{2} \right) }.
\ee
There is a non-zero $(p+2)$-form field strength $F^{(p+2)}$ corresponding to the
electric RR-potential $C^{(p+1)}$ of $N$ D$p$-branes. 
Here we rather characterize the background by dual magnetic $(8-p)$-form
field strength $F^{(8-p)}$ as
\be
F^{(8-p)}_{m_1m_2\cdots m_{8-p}} = - \epsilon_{m_1\cdots m_{8-p}n}
\partial_n H .
\ee
We see that infinite series of non-abelian Taylor expansion is needed 
if we write down the Chern-Simons term by using 
this expression of $F^{(8-p)}$ and the non-commutative coordinates $X^i$.
Fortunately, we find an appropriate coordinate system 
for which $F^{(8-p)}$ becomes constant.
The new coordinates $\{ r,\phi, z_{p+3},\cdots, z_9\}$ are given from 
$\{x^{p+1},\cdots,x^9\}$ by the relation
\bea
x_{p+1} &=& r \sqrt{1-(z_{p+3}{}^2+\cdots + z_9{}^2)} \cos\phi ,
\nonumber\\
x_{p+2} &=& r \sqrt{1-(z_{p+3}{}^2+\cdots + z_9{}^2)} \sin\phi ,
\nonumber\\
x_{p+3} &=& r z_{p+3} ,
\nonumber\\
\vdots \; &=& \; \vdots 
\nonumber\\
x_9 &=& r z_{9} .
\eea
The corresponding metric component is rewritten as
\bea
\delta_{mn}dx^{m} dx^{n} & = & dr^2 + r^2 [1-(z_{p+3}{}^2+\cdots + z_9{}^2)] d\phi^2
+r^2 (d z_{p+3}{}^2+\cdots +d z_9{}^2)
\nonumber\\
&  & + \frac{r^2}{1-(z_{p+3}{}^2+\cdots +z_9{}^2)} 
(z_{p+3} d z_{p+3}+\cdots + z_9 d z_9)^2 \; .
\label{metppart}
\eea
The explicit form of $F^{(8-p)}$ is given as
\be
F^{(8-p)}_{\mu_1\mu_2\cdots \mu_{8-p}}=
\left\{
\begin{array}{@{\,}cl}
-\frac{2\pi}{T_{6-p}V_{8-p}}N \epsilon_{\mu_1,\mu_2\cdots, \mu_{8-p}}
&  \mbox{for $\mu_1,\mu_2\cdots, \mu_{8-p} 
\in \{\phi, z_{p+3},\cdots, z_9\}$}\\
0 & \mbox{otherwise}
\end{array} 
\right.
\ee
where $\epsilon_{\phi z_{p+3}\cdots z_9}=1$. 
By using this expression,
the Chern-Simons action can be represented unambiguously
without using non-abelian Taylor expansion for each $p$.

In order to give the Born-Infeld action of $N_0$ coincident D0-branes
explicitly,
we first assume that the coordinates $r$ and $\phi$
remain to be commutative fields 
on the D0-branes.
This restriction is necessary for the action to be tractable and 
also can be understood by the fact that the meaning of matrix generalization 
of a radial or an angular coordinate is less clear than that of 
flat-like coordinate $Z^{\hat{i}}$ or $X^a$.
Furthermore, we assume $X^a=0$ ($a=1,\cdots,p$) for simplification.
The other $7-p$ transverse coordinates 
become adjoint scalar fields $Z^{p+3},\cdots,Z^{9}$ on the branes.
Then, the action is written as 
\be
{\cal L}_{BI} =
-T_0 {\rm Tr} \left(
H^{\frac {p-4} 4 }\sqrt{\left[1-H\dot{r}^2 -H r^2 
(1-Z^{\hat{i}}Z^{\hat{i}})\dot{\phi}^2- H^{\frac 1 2}
(Q^{-1})^{\hat{j}}{}_{\hat{k}}\dot{Z}^{\hat{k}}\dot{Z}^{\hat{i}}G_{\hat{i}\hat{j}}
\right]det (Q^{\hat{i}}{}_{\hat{j}})
}
\right)
\label{BIpr}
\ee
where $\hat{i}=p+3,\cdots,9$.
We expand the action in terms of $[Z^{\hat{i}},Z^{\hat{j}}]$
or $\dot{Z}^{\hat{i}}$ and take the leading contribution: 
\bea
\sqrt{det (Q^{\hat{i}}{}_{\hat{j}})} &=& 1- \frac {1}{4 \lambda^2}
 H r^4 ([Z^{\hat{i}},Z^{\hat{j}}]^2) + \cdots \, ,
\\
(Q^{-1})^{\hat{j}}{}_{\hat{k}}\dot{Z}^{\hat{i}}\dot{Z}^{\hat{k}}
&=&
\dot{Z}^{\hat{j}}\dot{Z}^{\hat{i}}+\cdots\, 
.
\eea
We also consider the non-relativistic limit.
Then the action becomes
\bea
{\cal L}_{BI} 
&=& 
-N_0 T_0 H^{\frac {p-4} 4 } \left\{
1-\frac{1}{2 N_0}H^{\frac 1 2} 
{\rm Tr}(\dot{Z}^{\hat{i}}\dot{Z}^{\hat{j}}G_{\hat{i}\hat{j}}) 
-\frac 1 2 H\dot{r}^2 - \frac {1}{4 N_0 \lambda^2}
 H r^4 {\rm Tr} ([Z^{\hat{i}},Z^{\hat{j}}]^2)
\right.
\nonumber\\
& & \left. -\frac 1 2 H r^2 
\left[1-\frac 1 {N_0} {\rm Tr}(Z^{\hat{i}}Z^{\hat{i}} )\right] 
\dot{\phi}^2 
\right\}\,.
\label{BIpe}
\eea

\subsection{$p=4$}
We explicitly consider the $p=4$ case. 
The background four-form field strength is 
\be
F^{(4)}_{\phi z_7 z_8 z_9} = -\frac{2\pi}{T_{2}V_{4}}N .
\ee
By substituting this into eq.(\ref{CS3}) and use 
partial derivative operation properly, the 
Chern-Simons part of the Lagrangian is given as
\be
{\cal L}_{CS} = -\frac{i}{2} N 
{\rm Tr} (Z^{\hat{i}} Z^{\hat{j}} Z^{\hat{k}})\dot{\phi} 
\epsilon_{{\hat{i}} {\hat{j}} {\hat{k}}}.
\ee
Combining this with eq.(\ref{BIpe}),
the total Lagrangian becomes 
\bea
{\cal L}&=& -N_0 T_0 +
\frac{1}{2} N_0 T_0 H \dot{r}^2 +
\frac{1}{2} N_0 T_0 H r^2
\left[1-\frac 1 {N_0} {\rm Tr}(Z^{\hat{i}}Z^{\hat{i}})\right] 
\dot{\phi}^2 + 
\frac{T_0}{2}H^{\frac 1 2} {\rm Tr}
(\dot{Z}^{\hat{i}}\dot{Z}^{\hat{j}}G_{\hat{i}\hat{j}})
\nonumber\\
&& + 
\frac{T_0}{4 \lambda^2}H r^4
{\rm Tr} ([Z^{\hat{i}},Z^{\hat{j}}]^2)
-\frac{i}{2} N 
{\rm Tr} (Z^{\hat{i}} Z^{\hat{j}} Z^{\hat{k}})\dot{\phi} 
\epsilon_{{\hat{i}}{\hat{j}}{\hat{k}}}\, .
\label{N0D0p4}
\eea
Note that $H=1+N\lambda/2T_0 r^3$ from  eq.(\ref{condNk}).

Equations of motion with respect to $Z^{\hat{i}}$ and $\phi$ are 
obtained as
\be
T_0
\frac{d}{dt} \left(\dot{Z}^{\hat{j}}G_{\hat{i}\hat{j}}H^{\frac 1 2} \right)
+ \frac{r N}{\lambda}\left[
\frac{\lambda T_0}{N} r H \dot{\phi}^2
Z^{\hat{i}}
- \frac{T_0}{\lambda N}r^3 H [ [ Z^{\hat{j}},Z^{\hat{i}}], Z^{\hat{j}}] +
\frac {3}{4} i \left(\frac{\lambda}{r}\dot{\phi}\right)
[Z^{\hat{j}},Z^{\hat{k}}]\epsilon_{\hat{i}\hat{j}\hat{k}}
\right]=0,
\label{eomZp4}
\ee
\be
P_{\phi} \equiv
N_0 T_0 H r^2 \dot{\phi}
\left[1-\frac 1 {N_0} {\rm Tr}(Z^{\hat{i}}Z^{\hat{i}})\right]
-\frac{i}{2}N {\rm Tr} (Z^{\hat{i}} Z^{\hat{j}} Z^{\hat{k}}) 
\epsilon_{{\hat{i}}{\hat{j}}{\hat{k}}} =constant.
\label{eomphip4}
\ee
%
We assume that each $Z^{\hat{i}}$ is a constant matrix.
There is a pointlike solution satisfying $Z^{\hat{i}}=0$ and
$Hr^2 \dot{\phi}\!=\!\mbox{constant}$.
Motion along $r$ is determined by solving equation of motion with respect to $r$.
There is no other solution found in general in this class.

If we restrict the geometry to be near horizon region of D4-branes, situation
becomes more interesting.
In this case, since $H=N \lambda/ 2 T_0 r^3$, 
$\dot{\phi}/r$ is restricted to be constant from eq.(\ref{eomphip4})
if we assume $Z^{\hat{i}}$ to be a constant matrix. 
Then eq.(\ref{eomZp4}) is explicitly solved by
\be
Z^{\hat{i}} =w \alpha^{\hat{i}} 
\label{Zw}
\ee
with
\be
w = 0, \quad \frac{\lambda \dot{\phi}}{2 r},\quad {\rm or} \quad
\frac{\lambda \dot{\phi}}{4 r} .
\label{sols2}
\ee
Here 
$\alpha^{\hat{i}}$ ($\hat{i}=7,8,9$)
 is $N_0$ dimensional representation of SU$(2)$:
\be
[\alpha^i,\alpha^j]=2i\epsilon_{ijk}\alpha^k . 
\label{su2alpha}
\ee
For each constant $w(\ne 0)$, the solution represents a fuzzy two-sphere of
effective radius 
measured by $z^{\hat{i}}$ \cite{My} 
\be
r_z=w N_0 \sqrt{1-\frac{1}{N_0{}^2}} \sim w N_0
\ee
where $r_z{}^2\equiv \frac{1}{N_0}{\rm Tr}(Z^{\hat{i}}Z^{\hat{i}})$.
Behavior of $r$ is determined by solving equation of motion for $r$.
To see the energy of the solutions, we calculate the
Hamiltonian of this system for $Z^{\hat{i}}$ satisfying eq.(\ref{Zw}) :
\be
{\cal H}=T_0 N_0 + \frac{r^3}{\lambda N N_0} P_r{}^2
+ \frac{r}{\lambda N N_0}
\left\{
P_{\phi}{}^2+\frac{w^2 (N_0{}^2-1)}{1-w^2(N_0{}^2-1)}
(P_{\phi}-w N N_0)^2
\right\}.
\ee
For a given $P_{\phi}$ and a particular motion along $r$,
Hamiltonian has two degenerate minima 
\be
w=0\; , \frac{P_{\phi}}{NN_0}.
\ee
Note that the relation 
\be
P_{\phi}=\frac{\lambda N N_0}{2r}\dot{\phi}
\ee
is satisfied for both values of $w$ and they correspond to 
two of the solutions eq.(\ref{sols2}): $w = 0, \, \frac{\lambda \dot{\phi}}{2 r}$. 
The other solution is at unstable point of the energy which is 
double-well form as in the case of giant gravitons \cite{MST,DTV,GMT,HHI}.
To summarize, 
we have two configurations of $N_0$ D0-branes:
one is point-like and the other is fuzzy $S^2$ with the radius 
$r_{z}=w N_0$ for $N_0\rightarrow \infty$.
Definition of the coordinates $z_{\hat{i}}$ restricts the radius as
$0\le r_{z}\le 1$.
Then for the expanding configuration, $P_\phi$ cannot exceed $N$:
\be
P_{\phi}\le N .
\ee 

\medskip

To discuss the problem beyond the expansion eq.(\ref{BIpe}), 
we substitute eq.(\ref{Zw}) with an arbitrary $w=w(t)$ into 
the original Born-Infeld action eq.(\ref{BIpr}). 
The resulting action has a simple form 
\bea
S_{BI}\left(Z^{\hat{i}}\!=\!w\alpha^{\hat{i}}\right) &=&
-T_0 N_0 \int dt
\sqrt{1-H\dot{r}^2 -H r^2 
(1- r_z{}^2)\dot{\phi}^2 -Hr^2\frac{1}{1\!-\!r_z{}^2}\dot{r_z}^2}
\nonumber\\
&&\hspace*{1cm} \times \sqrt{1+\frac{4 H r^4 r_z{}^4}{\lambda^2(N_0{}^2-1)}}\, .
\label{BIp4zw}
\eea
Note that here the only approximation we use is that 
we take the leading contribution of 
$(Q^{-1})^{\hat{i}}{}_{\hat{k}}= \delta^{\hat{i}}{}_{\hat{k}}+\cdots$. 
Remarkably, if we take $N_0\rightarrow \infty $ limit,
this action precisely coincides with that of 
a spherical D2-brane with $N_0$ D0-branes bound on it.
This can be seen by considering a spherical D2-brane of 
worldvolume $(t,\theta,\psi)$ on which U(1) field strength 
\be
F_{\theta\psi}=\frac{N_0}{2} \sin \theta 
\ee 
describing $N_0$ D0-branes exists.
Here 
\be
dz_{\hat{i}}dz_{\hat{i}}= dr_z{}^2 +r_z{}^2(d\theta^2+\sin^2\theta d\psi^2) .
\ee
Assuming that $r=r(t)$ and 
$r_z(\equiv \sqrt{z_{\hat{i}}z_{\hat{i}} })=r_z(t)$,
the D2-brane action is \cite{DTV}
\bea
S^{D2} &=& -T_2\int dt d\theta d\psi e^{-\phi}
\sqrt{-det(P[G_{ab}+\lambda F_{ab}])} +T_2\int P[{C^{(3)}}]
\nonumber\\
&=& -4\pi T_2\!\int dt \sqrt{1-\!H \dot{r}^2 
\!-Hr^2(1-r_z{}^2)\dot{\phi}^2
-H r^2 \frac{1}{1-r_z{}^2} \dot{r_z}^2
}
\sqrt{r^4 r_z{}^4 H +\frac{N_0{}^2 \lambda^2}{4}}
\nonumber\\
& &+N\!\int dt r_z{}^3 \dot{\phi}
\eea
which is equivalent to eq.(\ref{BIp4zw}) up to $1/N_0{}^2$ correction.

\subsection{$p=2$}
Next we consider $N_0$ D0-branes in the geometry of $N$ 
D2-branes.
D0-branes couple to the background six-form field strength 
\be
F^{(6)}_{\phi z_5 z_6 z_7 z_8 z_9} = -\frac{2\pi}{T_{4}V_{6}}N .
\ee
Then from eq.(\ref{CS5}),
Chern-Simons term is calculated as
\be
{\cal L}_{CS} = -\frac{3}{4} N 
{\rm Tr} (Z^{\hat{i}} Z^{\hat{j}} Z^{\hat{k}} Z^{\hat{l}} Z^{\hat{m}})\dot{\phi} 
\epsilon_{{\hat{i}} {\hat{j}} {\hat{k}}{\hat{l}}{\hat{m}}}.
\ee
Combining with ${\cal L}_{BI}$ of eq.(\ref{BIpe}), we have
\bea
{\cal L}&=& -N_0 T_0 H^{-\frac 1 2}\left\{
1-\frac{1}{2} H \dot{r}^2 -
\frac {1}{2}H r^2 
\left[1-\frac 1 {N_0} {\rm Tr}(Z^{\hat{i}}Z^{\hat{i}})\right] 
\dot{\phi}^2 - \frac{1}{2 N_0} H^{\frac 1 2}
{\rm Tr}(\dot{Z}^{\hat{i}}\dot{Z}^{\hat{j}} G_{\hat{i}\hat{j}})
\right.
\nonumber\\
&& \left. -\frac {1}{4 N_0 \lambda^2}H r^4 {\rm Tr} ([Z^{\hat{i}},Z^{\hat{j}}]^2)
\right\}-
\frac{3}{4} N 
{\rm Tr} (Z^{\hat{i}} Z^{\hat{j}} Z^{\hat{k}} Z^{\hat{l}} Z^{\hat{m}})\dot{\phi} 
\epsilon_{{\hat{i}} {\hat{j}} {\hat{k}}{\hat{l}}{\hat{m}}}
\label{N0D0p2}
\eea
where $H=1+\frac{k}{r^5}$ with $k=3N\lambda^2/2 T_0$.
If we assume $Z^{\hat{i}}=0$, then the motion of $\phi$ and $r$ is determined from
eq.(\ref{N0D0p2}).
As in the previous subsection, we expect to find a constant non-commutative
$Z^{\hat{i}}\ne 0$ which solves equations of motion and has the same motion as for the commutative solution.
In this case, we cannot obtain such a solution
even in the near horizon limit.

On the other hand, we can find a solution 
with $Z^{\hat{i}}= Z^{\hat{i}} (t)$ though such a configuration is 
not related to the point-like configuration. 
In particular, we will concentrate on a configuration 
which is related to a spherical D4-brane of radius $r_z=r_z(t)$.
For this aim, we need to represent a fuzzy four-sphere 
of radius $r_z$ by $Z^{\hat{i}}$:
\be
r_z{}^2 = \frac{1}{N_0}{\rm Tr} (Z^{\hat{i}} Z^{\hat{i}}) .
\ee
This has been constructed in the context of longitudinal 5-brane of 
Matrix theory \cite{BFSS} 
as in ref.\cite{CLT,GKP}.
According to ref.\cite{CLT}, fuzzy $S^4$ is represented by
$N_0 \times N_0$ matrices
$G_i^{(n)}$ ($i = 1, \cdots , 5 $) that are given by
$n$-fold symmetric product of SO$(5)$ $4 \times 4$ gamma matrices $\gamma^i$ :
\be
G_i^{(n)} = (\gamma_i\otimes{\bf 1}\otimes \cdots \otimes {\bf 1}
+{\bf 1}\otimes \gamma_i \otimes \cdots \otimes {\bf 1}
+\cdots +
{\bf 1}\otimes \cdots \otimes {\bf 1}\otimes \gamma_i
)_{sym}\, .
\ee 
Here $sym$ denotes that 
the tensor product space is restricted to be 
completely symmetric and the 
dimension of the matrices $G_i^{(n)}$ is 
$(n+3)!/3!n!=(n+1)(n+2)(n+3)/6$.
This means that the
number of D0-branes is restricted to $N_0=(n+1)(n+2)(n+3)/6$.
Let us take 
\be
Z^{\hat{i}} =w G_{\hat{i}-4}^{(n)}.
\label{ZiGi}
\ee
Then from the property of $ G_i^{(n)}$, we find that \cite{CLT}
\be
r_z{}^2 \equiv \sum_{i}(Z^{\hat{i}}){}^2 =w^2 n^2 \left(
1+\frac{4}{n} \right) {\bf 1}_{N_0}.
\ee
This means that the radius of non-commutative $S^4$ measured by
$Z^{\hat{i}}$ is $r_z=wn$ for large $n$.
There are some useful relations:
\bea
\epsilon_{\hat{i}\hat{j}\hat{k}\hat{l}\hat{m}} 
Z^{\hat{i}} Z^{\hat{j}} Z^{\hat{k}} Z^{\hat{l}}  
& = & (8n+16)w^3 Z^{\hat{m}} ,
\\
{\rm Tr} [Z^{\hat{i}}, Z^{\hat{j}}]^2 
& = & -16 n(n+4)w^4 N_0 ,
\\
{[} [Z^{\hat{j}} , Z^{\hat{i}}] , Z^{\hat{j}}] 
& = & -16 w^2 Z^{\hat{i}} .
\eea
We take an ansatz
\be
Z^{\hat{i}}= w(t) G_{\hat{i}}^{(n)} .
\ee
Substitution of this in the action eq.(\ref{N0D0p2}) 
yields 
\bea
{\cal L}&=& -N_0 T_0 H^{-\frac{1}{2}}\left[
1-\frac{1}{2} H \dot{r}^2 -\frac {1}{2}H r^2 (1-r_z{}^2 )\dot{\phi}^2- 
\frac{1}{2 } H \frac{r^2}{1-r_z{}^2} 
\dot{r}_z{}^2 +\frac {2}{3 N_0 \lambda^2} H r^4 r_z{}^4
\right]
\nonumber\\
&&
-n N r_z{}^5\dot{\phi}.
\label{D0p4}
\eea
in the  $N_0 (\sim n^3/6) \rightarrow \infty$ limit.
By solving equations of motion, we can determine the classical 
motion of the fuzzy 4-spherical D0-branes.

\bigskip

Next, we will see from the dual D4-brane picture that this system 
is corresponding to $n$ coincident spherical 
D4-branes with $N_0$ D0-branes on them
when $N_0 \gg 1$.
To show this,
we put $n$ coincident spherical D4-branes of radius $r_z$
in the same background.
We transform coordinate 
from $\{z_5,\cdots,z_9 \}$ to
$\{r_z,\theta_1,\theta_2,\theta_3,\psi\}$ as
\bea
z_5&=&r_z \cos\theta_1,\; z_6=r_z\sin\theta_1 \cos\theta_2,\; 
z_7=r_z \sin\theta_1 \sin\theta_2 \cos\theta_3,
\nonumber\\
z_8&=&r_z \sin\theta_1 \sin\theta_2 \sin\theta_3 \cos \psi,\;
z_9=r_z \sin\theta_1 \sin\theta_2 \sin\theta_3 \sin \psi. 
\eea
We assume that the worldvolume of spherical D4-branes is labeled by 
$\{t, \theta_i, \psi\}$ and the transverse 
U($n$) scalar fields are all commutative with
$x^1=x^2=0$, $r=r(t)$, $\phi=\phi(t)$ and $r_z=r_z(t)$. 
Background five-form potential $C^{(5)}$ is written by new coordinates
by choosing a gauge :
\be
C^{(5)}_{\phi \theta_1\theta_2\theta_3\psi } 
= \frac{2 \pi N}{5 T_4 V_6} r_z{}^5
\sin^3\theta_1 \sin^2\theta_2 \sin\theta_3 .
\ee
The Chern-Simons action for $n$ coincident (anti) D4-branes is%
\footnote{
We take anti D4-branes to obtain the same sign as eq.(\ref{D0p4}).
}
\bea
S_{CS}^{n\, D4} & = & - n T_4 \int dt \left(
\frac{2\pi N}{5T_4 V_6} r_z{}^5 \right)
\int_{S^4} d\theta_1 d\theta_2 d\theta_3 d\psi
\dot{\phi}\sin^3\theta_1 \sin^2\theta_2 \sin\theta_3
\nonumber\\
& = & - n N \int dt \dot{\phi} r_z{}^5 .
\eea
Next we consider Born-Infeld action. 
The self-dual field strength $F_{ab}$ satisfying 
\be
F=\ast_{4}F ,\quad
\frac{1}{8\pi^2}\int_{S^4}{\rm Tr}_n F\wedge F =N_0 
\label{SDFab}
\ee
is on the D4-branes.
This describes that $N_0$ D0-branes are bound on D4-branes.
The normalization is confirmed by the fact that the coupling 
$C^{(0)}\wedge F \wedge F$ in $S_{CS}$ reproduces the coupling to
$N_0$ D0-branes as
\be
 T_4 \int_{S^4 \times \{ t \} } \frac{\lambda^2}{2}
{\rm Tr}_n (C^{(0)}\wedge F \wedge F)
= T_0 N_0 \int_{\{ t  \} } C^{(0)} \; .
\ee
Then the action is
\bea
S_{BI}^{n\, D4} &=& -T_4 \int dt \int_{S^4} e^{-\phi} {\rm Tr}_n
\sqrt{-\det (P[G_{ab}+\lambda F_{ab}])}
\nonumber\\
&=& -T_4 \int dt\, H^{-\frac{1}{2}}
\sqrt{1- H\dot{r}^2 -H \frac{r^2}{1-r_z{}^2}\dot{r_z}^2 -
H(1-r_z^2)r^2 \dot{\phi}^2}
\nonumber\\
&& \quad \times \int_{S^4} {\cal \epsilon}_{4} 
\left(n H r^4 r_z{}^4 + \frac{\lambda^2}{4}{\rm Tr}_n F_{ab}F^{ab}
\right)
\nonumber\\
& = & - T_4 \int dt \, H^{-\frac{1}{2}}
\sqrt{1- H\dot{r}^2 -H \frac{r^2}{1-r_z{}^2}\dot{r_z}^2 -
H(1-r_z^2)r^2 \dot{\phi}^2}
\nonumber\\
&& \quad \times
\left(n V_4 H r^4 r_z{}^4 + 4 \pi^2 N_0 \lambda^2 \right)
\eea
where 
${\cal \epsilon}_{4}=\sin^3\theta_1 \sin^2\theta_2 \sin\theta_3 $
is volume element of $S^4$.
If there is no D0-brane on the D4-branes, it is known that there exist
two degenerate
configurations $r_z=0$ and $r_z{}^3=P_{\phi}/N$
for a constant $P_{\phi}$ \cite{DTV}.
Existence of D0-brane charge makes the expanding configuration unstable.
Note that this property is different from the case of a spherical D2-brane
in the D4-brane background
where the D0-brane charge on a spherical D2-brane does not affect the stability 
of the radius $r_z$ of the expanding configuration.

At this point, we see that there is a non-trivial correspondence
between this picture from spherical D4-branes and that from 
non-commutative D0-branes.
Expansion of $S_{BI}^{n\,D4}+S_{CS}^{n\,D4}$
includes all terms of D0-brane action eq.(\ref{D0p4})
and the coefficients all agree with each other.
Missing terms in eq.(\ref{D0p4}) would be obtained if we consider
beyond the leading contribution of $[Z^{\hat{i}},Z^{\hat{j}}]$ in 
$det(Q^{\hat{i}}{}_{\hat{j}})$.
This property supports the fact that a fuzzy 4-sphere configuration of D0-branes 
in the D2-brane background certainly represents a bound state of $n$ coincident
spherical D4-branes and $N_0$ D0-branes on them.

\section{Summary and Discussion}
We have investigated the non-abelian worldvolume effective action 
of coincident D0-branes
in general backgrounds.
In particular, we considered the action in the
D$p$-brane geometry with $p=2$ or $p=4$. 
For such a spacetime, we have shown that coupling of $N_0$
D0-branes with background $(8-p)$-form field strength in the 
Chern-Simons action can be written explicitly if we 
choose a particular coordinate system.
By using the expression, the non-abelian action of D0-branes can 
be analyzed as a microscopic theory.
Especially for $p=4$ or $p=2$, it was shown that fuzzy $(6-p)$-sphere
configuration is certainly regarded as a bound state of $D(6-p)$-branes 
and D0-branes.

Moreover, if we consider near horizon geometry of 
$p=4$, we explicitly solved the equations of motion of the 
action written in particular coordinates and have found two solutions 
that have same properties, e.g., momentum and charge.
One is point-like and the other is expanding into a form of a spherical D2-brane.
It seems that
the two degenerate configurations of D0-branes 
may be BPS states preserving some of the supersymmetries.
The precise relation between the non-abelian D0-branes with 
$X^{\hat{i}}=w\alpha^{\hat{i}}$ and the dual spherical D2-brane on which U(1)
field strength lives has been clarified.
Since there are two degenerate configurations with 
the same quantum numbers in the system,
it is reminiscent of giant gravitons \cite{MST,GMT,HHI}.
On the other hand, for $p=2$, we could not find the corresponding structure 
of expanding brane configurations as in $p=4$.
This may be related to the fact that no supersymmetries remain if 
$N_0$ D0-branes and $N$ D2-branes both exist.

\bigskip

Now we briefly discuss the relation between 
expanding brane configurations via Myers effect  
and 11-dimensional giant gravitons.
Giant graviton (or dual giant graviton)
in $AdS_{m}\times S^n$ with $(m,n)=(7,4)$ or $(4,7)$
is known as a spherical M$(n-2)$-brane (or M$(m-2)$-brane) shaped object
expanding into $S^{n-2} \subset S^n$ (or $S^{m-2} \subset S^m$)
with non-zero angular momentum $P_{\phi}$ along $S^{n}$ \cite{MST,GMT,HHI}.
We cannot obtain a non-singular 10-dimensional configuration by explicit
compactification of these systems.
However, as was discussed in ref.\cite{DTV,Lo}, 
there are some D0-brane systems 
which are considered as ten-dimensional counterparts 
of giant gravitons in a sense that each framework resembles
each other
in the mechanism of expansion of D0-branes or gravitons.
For example, compactification of the system of 
dual giant graviton expanding into 
spherical M2-brane in $AdS_{4}\times S^7$ along $\phi$ direction
corresponds to dielectric D0-branes expanding into spherical D2-brane 
under the RR field $F^{(4)}_{0123}\ne 0$.
11-dimensional momentum $P_{\phi}$ along $\phi$ changes into D0-branes 
by the relation $P_{\phi}=N_0/gl_s$.
Similarly, 10-dimensional counterpart of 
giant graviton expanding into $S^2(\subset S^4)$ in
$AdS_{7}\times S^4$ is $N_0$ D0-branes expanding into 
spherical D2-brane by the background NS-NS field strength $H_{789}$.
This relation is obtained when we presumably set that 
the 11-dimensional direction is along $\phi$.
We can relate this giant graviton system to that of D0-branes expanding
via magnetic moment effect under $F^{(4)}_{6789}\ne 0$.
To see this, we first transform coordinates of $AdS_7$ by using 
a particular isometry of the Anti de Sitter space.
With the transformation of time coordinate $t\rightarrow t'$, 
angular momentum changes as $P_{\phi}\rightarrow P'_{\phi} + P'_{\psi'}$ 
for some of the new coordinates $\psi'\in AdS_7$. 
Compactification of the system along $\psi'$ indicates the configuration
of D0-branes expanding via magnetic moment effect 
since $P'_{\phi}$ remains to be angular momentum and $P_{\psi'}$ changes to 
D0-brane charges.

Next, consider (dual) giant gravitons expanding into spherical M5-branes.
We expect that there is a similar relation between these configurations
and expanding D0-brane configurations in 10-dimensions.
If there exist corresponding configurations in 10-dmensions,
coincident D0-branes must couple to
$\ast H^{(3)}{}\!_{\mu ijklmn}$ and $F^{(6)}_{ijklmn}$ by 
\be
{\rm Tr}(\ast H^{(3)}{}\!_{\mu ijklmn}
\dot{X}^{\mu}X^{i} X^{j} X^{k} X^{l} X^{m} X^{n})
\ee
and 
\be
{\rm Tr}(F^{(6)}_{ijklmn}X^{i} X^{j} X^{k} X^{l} X^{m} X^{n})
\ee
respectively.
We see that the action eqs.(\ref{MpBI}) and (\ref{MpCS}) 
does not have corresponding terms.
They may appear in the non-perturbative action of D0-branes.
In practice, it was discussed in ref.\cite{Lo} that similar terms 
can be considered at strong coupling.
Note that D0-branes would expand into spherical NS5-branes and not into D4-branes
by these couplings.

\paragraph{Acknowledgments} The author would like to thank S. R. Das, 
Y. Imamura, M. Kato, Y. Matsuo, and especially 
Y. Sekino for discussions and comments.


\end{document}